\documentclass[aps,prb,twocolumn,superscriptaddress,showpacs,floatfix]{revtex4-2}
\usepackage{fullpage}
\usepackage{mathrsfs} \usepackage{amssymb} \usepackage{graphicx,color} \usepackage{bbm} \usepackage{multirow}
\usepackage{bm} \usepackage{mathrsfs} \usepackage{commath} \usepackage{stmaryrd} \usepackage{color}
\usepackage[sort&compress]{natbib}
\usepackage{subfigure}
\usepackage{comment}
\usepackage{inputenc}
\usepackage{ulem}
\usepackage{amsmath}
\usepackage{amssymb}
\usepackage{bm}
\usepackage{graphicx}
\usepackage{float}

\usepackage{hyperref}
\hypersetup{
  colorlinks   = true,    
  urlcolor     = blue,    
  linkcolor    = blue,    
  citecolor    = blue     
}

\newcommand{\beq}{\begin{equation}}
\newcommand{\eeq}{\end{equation}}

\graphicspath{ {figures/} }
\begin{document}

\title{Localization induced by spatially uncorrelated subohmic baths in one dimension}

\author{Saptarshi Majumdar}
\affiliation{Universit\'{e} Paris Saclay, CNRS,LPTMS, 91405, Orsay, France}

\author{Laura Foini}
\affiliation{IPhT, CNRS, CEA, Universit\'{e} Paris Saclay, 91191 Gif-sur-Yvette, France}

\author{Thierry Giamarchi}
\affiliation{Department of Quantum Matter Physics, University of Geneva, 24 Quai Ernest-Ansermet, CH-1211 Geneva, Switzerland}

\author{Alberto Rosso}
\affiliation{Universit\'{e} Paris Saclay, CNRS,LPTMS, 91405, Orsay, France}

\begin{abstract}
We study an incommensurate XXZ spin chain coupled to a collection of local harmonic baths. At zero temperature, by varying the strength of the coupling to the bath the chain undergoes a quantum phase transition between a Luttinger liquid phase and a spin density wave (SDW).
As opposed to the standard mechanism, the SDW emerges in the absence of the opening of a gap, but it is due to ``fractional excitations" induced by the bath. We also show, by computing the DC conductivity, that the system is insulating in the presence of a subohmic bath. We interpret this phenomenon as localization induced by the bath \`a la Caldeira and Leggett.
\end{abstract}

\date{\today}

\maketitle

\section{Introduction}
\begin{figure*}[t!]
\centering
\includegraphics[width=1\linewidth, clip, trim=40 10 30 100]{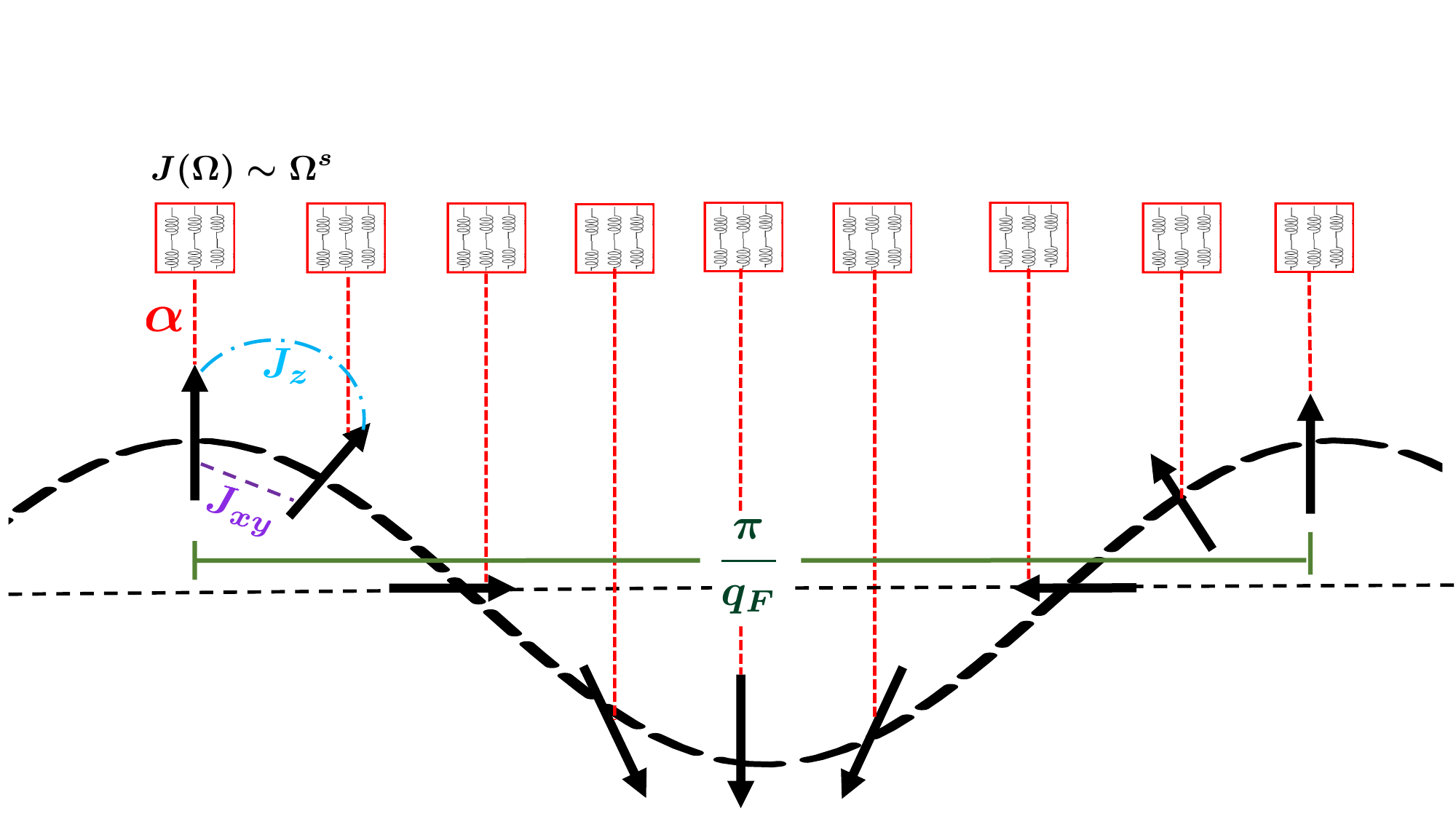}
\caption{Schematic diagram of a one-dimensional quantum XXZ spin chain coupled with local dissipative baths. $J_{xy}$ denotes the hopping energy and $J_z$ is the interaction between the two nearest neighbour spins. The baths are characterized by their spectral function $J(\Omega) \sim \alpha \Omega^s$. At zero temperature, the baths induce an SDW phase with periodicity $\pi/q_F$, where $q_F$ is the Fermi-momentum related to the magnetization of the chain (see text).}
\label{Fig_SDW}
\end{figure*}
Open quantum systems, namely systems coupled with external degrees of freedom, are often studied in order to understand the phenomenon of decoherence and the emergence of classical laws from a quantum mechanical description. A common setup is to consider the Markovian dynamics of quantum systems subject to repeated measurements \cite{WayneZeno,SudarshanZeno}. One of the most intriguing results is the possibility to observe a phase transition in the behavior of the quantum trajectories. The transition is controlled by the measurement rate: For a low rate the entanglement grows linearly in time while at a high measurement rate, it saturates at a finite value \cite{CircuitZeno1,CircuitZeno2,CircuitZeno3,CircuitZeno4,CircuitZeno5,CircuitZeno6,CircuitZeno7,CircuitZeno8,CircuitZeno9,CircuitZeno10}. Another important setup is to consider the effect of a thermal bath on the system. Following the pioneering works \cite{Chakravarty,leggettchakravarty,Bray&Moore,Schmid}, we expect that a slow bath (i.e. subohmic and ohmic) can induce localization in simple systems, such as  a particle or a spin. Note that this dynamical transition cannot be described by a Lindblad equation \cite{Thibaud}. Indeed, in order to capture this localization phenomenon it is crucial to relax the Markovian assumption which is behind the Lindblad equation. Moreover from  several variational studies of the ground state of the spin-boson model (namely, the Caldeira Leggett model for a single spin), a genuine thermodynamic transition has been shown to exist for strongly coupled subohmic bath \cite{Subohmic_transition1, hur2009quantum}.   

In this work, we investigate the possibility of such non-Markovian transition in many body systems. In particular, we focus on a one-dimensional (macroscopic) interacting and incommensurate spin chain coupled to {\it local} baths of harmonic oscillators (fig. \ref{Fig_SDW}). This problem was studied in \cite{Sap1} with a special focus on the ohmic case. Here we generalize the study to the superohmic and subohmic case, with particular emphasis on the nature of the dissipative phase both for thermodynamic and transport properties. In particular, we show that the dissipative phase is an incommensurate spin density wave of period $\pi /q_F$, where $q_F$ is the Fermi momentum of the system. Unlike the Peierls scenario \cite{Peirels}, this spin density wave emerges in the absence of the opening of a gap, but it is due to ``fractional excitations" induced by the slow varying bath.
The spin density wave order is not only particular to subohmic baths, but also survives in the presence of superohmic baths described by an exponent $s<2$. However, for subohmic bath, i.e. $s<1$, the environment can  induce ``localization" with a gapless insulating phase. The nature and the details of these ``fractional" dissipative phases are derived by studying the bosonized action with a thorough variational approach and tested with respect to the exact action with  numerical simulations for the subohmic case ($s=0.5$).

The metal-insulator transition for subohmic baths is reminiscent of the (zero temperature) localization transition which occurs in interacting one-dimensional systems due to the presence of quenched disorder \cite{GiamarchiSchulz2,GiamarchiSchulz1}. Indeed, local baths can be thought of as spatially uncorrelated annealed disorder. In the dissipative phase, the degrees of freedom of the system and those of the bath optimize collectively to find a low energy configuration \cite{LauraKurchan_annealed}.   

We also describe the finite size and finite temperature effects. At finite temperature, the order parameter vanishes but the spin density wave can be observed from correlation functions below a length scale which grows as $\beta$, where $\beta$ is the inverse temperature of the system. For finite system size (and zero temperature) the order parameter vanishes for $s>1$ and one recovers the phase transition that occurs for the spin-boson model with for subohmic baths \cite{hur2009quantum}.

The manuscript is organized as follows: in Section \ref{Sec_model} we introduce the model. The analytical variational solution of the model is described in Section \ref{Sec_var_ansatz}. Section \ref{Sec_OP} consists of detailed discussions about the nature of the order parameter and the dissipative phase, followed by the comparison of the analytical solution obtained with the variational ansatz with exact numerical simulation; in Section \ref{Sec_NS}. In Section \ref{Sec_Cond},  we discuss the transport properties of the model, and in Section \ref{Sec_Concl} we conclude about the nature of the dissipative phase and the absence of linear response transport in the system.

\section{Model}\label{Sec_model}

We investigate the zero-temperature low-energy phase diagram of an incommensurate XXZ spin chain in the presence of local subohmic baths. The Hamiltonian of the system is given by: 
\begin{equation}
    \begin{split}
        H &= H_{\text{S}}+H_{\text{B}}+H_{\text{SB}} \\
        H_\text{S} &= \sum_{j=1}^L J_z \sigma_j^z \sigma_{j+1}^z +J_{xy} \left(\sigma_j^x\sigma_{j+1}^x + \sigma_j^y \sigma_{j+1}^y\right)+ h\sigma_j^z\\ 
        H_{\text{B}} &= \sum_{j k} \frac{P^2_{jk}}{2 m_k} + \frac{m_k \Omega_k^2}{2} X_{jk}^2 \\
        H_{\text{SB}} &= \sum_{j=1}^N \sigma_j^z \sum_k \lambda_k X_{jk}
    \end{split}
\end{equation}

The dissipative baths are characterized by their spectral function $J(\Omega)\equiv \frac{\pi}{2} \sum_k (\lambda_k^2/m_k \Omega_k) \delta(\Omega-\Omega_k) = \pi \alpha \Omega^s$. In one dimension, XXZ spin chain is a general description of an interacting many-body system as it can be mapped onto spinless Fermionic chain and hard-core Bosonic chain via Jordan-Wigner \cite{JWT} and Holstein-Primakoff transformation \cite{Holstein} respectively. Its phase diagram is well known; particularly, at zero temperature and in finite magnetization sector ($h \neq 0$), one can use bosonization to arrive at the so-called Luttinger Liquid (LL) action \cite{giamarchibook}:
\begin{equation}
    S_{\text{LL}}= \frac{1}{2 \pi K} \int dx d\tau \left[ \frac{1}{u} (\partial_{\tau} \phi(x,\tau))^2 + u(\partial_{x} \phi(x,\tau))^2 \right]
\end{equation}
 where $\phi(x,\tau)$ is a bosonic field defined in the two-dimensional space of position $x \in (0,L)$ and imaginary time $\tau \in (0,\beta)$, $\beta$ being the inverse temperature of the system. $u$ is the speed of sound, $K$ is called Luttinger parameter and depends on the values of $J_z$ and $J_{\text{xy}}$. The contribution coming from the magnetic field, given by $-\frac{h}{\pi} \int \partial_x \phi$ in the bosonic language, can be absorbed into the action by using a tilt transformation $\phi \to \phi - h K x/u$. In this case, the Fermi momentum of the system $q_F=\pi(1-(M/N))/2a$ is incommensurate with the lattice spacing, hence we refer to the system as 'incommensurate spin chain'. Here $N$ is the total number of spins, $M$ is the total magnetization of the chain and $a$ is the lattice spacing. This action is known to describe a metallic, perfectly conducting, and gapless phase.

To analyze the effect of the bath on the spin chain, we apply bosonization to map the $\sigma_j^z$ operator onto the bosonic fields $\phi$ \cite{giamarchibook}:
\begin{equation}
    {\sigma}^z(x) = \frac{1}{\pi}\left( -\nabla \phi + \frac{1}{a} \cos \left(2 \phi(x) - 2 q_F x \right) \right)
    \label{eq:sigma-boson}
\end{equation}
Then we integrate out the bath degrees of freedom to arrive at an effective field theory (more details can be found in Sec. III, \cite{Sap1}):
\begin{eqnarray}
     S_{\text{eff}} &=& S_{\text{LL}}+S_{\text{diss}} 
     \label{eq:action}
 \\
 S_{\text{diss}} &=& -\frac{\alpha}{4\pi^2} \int dx d\tau d\tau' \frac{\cos \left(2\left(\phi(x,\tau)-\phi(x,\tau') \right)\right)}{|\tau-\tau'|^{1+s}} \nonumber
\end{eqnarray}

The local dissipative baths introduce a long-range cosine potential acting only along the $\tau$ direction, which can break symmetry and induce phase transition on the existing LL phase \cite{Ruffo,Lobos_superconducting}. A similar problem but with a single degree of freedom (particle) was shown to lead to phase transitions as a function of the exponent $s$ \cite{ThierryLedoussal1, ThierryLedoussal2}. In the subsequent sections, we show that the ordered dissipative phase is described by an SDW of the form:
\begin{equation}
    \langle \sigma^z(x) \rangle = \sigma_0 + \sigma_1 \cos(2 q_F x)
    \label{eq:SDW}
\end{equation}
Here $\sigma_0$ is the magnetization per spin $\sigma_0 = M/N$, while $\sigma_1$ is the amplitude of the SDW, which is the order parameter of the transition.
\section{Variational ansatz}\label{Sec_var_ansatz}

The action from eq. (\ref{eq:action}) can't be exactly solved due to the presence of the cosine term. One can estimate the critical properties of the action using a perturbative RG method \cite{cazaillashort} (see also Appendix B). However, here we rely on the variational method \cite{feynman1998statistical} to describe the nature of the different phases:  We find the best quadratic action $S_{\text{var}} = \frac{1}{2 \pi \beta L} \sum_{q,\omega_n} \phi^*(q,\omega_n) G^{-1}_{\text{var}}(q,\omega_n) \phi(q,\omega_n)$ that describes the original action effectively at zero temperature. One can write the free energy of the original system as $F_{\text{eff}} = T \log Z_{\text{eff}}=F_0 - T \log \left[ \langle \exp(S_{\text{eff}} - S_{\text{var}})\rangle_{S_{\text{var}}} \right]$, where $F_0=-T \ln Z_{\text{var}}$, $Z_{\text{eff}}$ is the exact partition function of the action that one wants to study and T is the temperature of the system. Now, we define a variational free energy $F_{\text{var}} = -\frac{1}{\beta} \sum_{q,\omega_n} \log G(q,\omega_n) + \frac{1}{\beta} \langle S_{\text{eff}} -S_{\text{var}}\rangle_{S_{\text{var}}}$. Due to the inequality $\langle \exp(-(S_{\text{eff}} - S_{\text{var}})) \rangle > \exp(-\langle (S_{\text{eff}} - S_{\text{var}})\rangle)$, it can be easily observed that $F_{\text{var}} \geq F_{\text{eff}}$. Hence, we minimize $F_{\text{var}} $ with respect to the variational propagator by setting $\frac{\partial F_{\text{var}}}{\partial G_{\text{var}} }=0$ to obtain a quadratic propagator that describes the system effectively. Applying this protocol to the action eq. (\ref{eq:action}), we find a self-consistent equation for $G^{-1}_{\text{var}}$:
\begin{eqnarray}
&&G^{-1}_{\text{var}} =  \frac{1}{\pi K}\left(uq^2+\frac{\omega_n^2}{u} \right) +\frac{\alpha}{\pi^2}\int\limits_{\tau_c}^{\infty} d\tau \frac{1-\cos \omega_n \tau}{\tau^{s+1}}\nonumber\\
&&\times \exp\left(- \frac{4}{\pi^2} \int\limits_{0}^{\infty} dq' d\omega_{n'} \ G_{\text{var}}\left(1-\cos \omega_{n'}\tau \right)\right)
\label{eq:self_con}
\end{eqnarray}
Where $\tau_c$ is the time-scale after which the bath displays the power-law behavior. In the next two subsections, we describe the analytical solution of this self-consistent equation. In the third subsection, we provide numerical evidence that supports this solution.

\begin{figure*}[t!]
\centering
\includegraphics[width=1\linewidth, clip, trim=10 55 5 30]{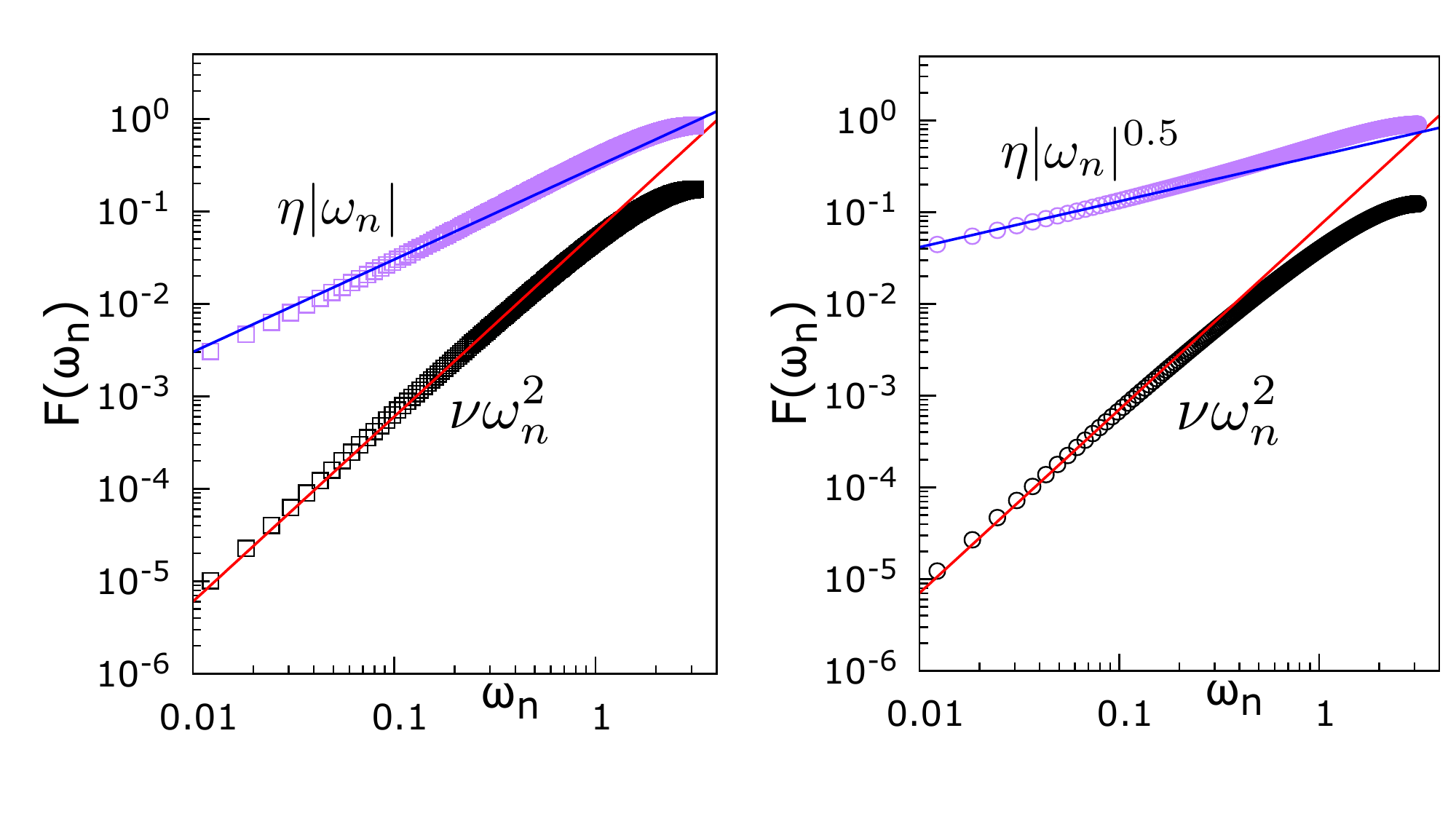}
\caption{$F(\omega_n)$ for ohmic ($s=1$, \textit{left}) and subohmic ($s=0.5$, \textit{right}) bath obtained by numerical solution of Eq. (\ref{Eq_num}) (with $\beta=1024$ and $\alpha=5$). In the dissipative phase, $ F(\omega_n)$ behaves as $0.301|\omega_n|$ (purple square points) for ohmic ($K=0.15$) and $0.415\sqrt{|\omega_n|}$ for (purple circular points) subohmic bath ($K=0.3$). In the LL phase, $F(\omega_n)=0.06\omega_n^2$ for ohmic bath (black square points) and $F(\omega_n)=0.054\omega_n^2$ for subohmic bath ($K=1$) (black circular points).}
\label{Fig_F_omega}
\end{figure*}

\subsection{Dissipative phase}

We first observe that the dissipative phase is gapless. Namely, for $q=\omega_n=0$, from eq. (\ref{eq:self_con}), we get $\Delta \equiv G^{-1}_{\text{var}}(q=0,\omega_n=0)=0$. Secondly, since $S_{\text{diss}}$ is invariant under a tilt transformation $\phi \to \phi -\frac{h \phi x}{\pi}$, the susceptibility is not affected by the potential, namely $\chi = \lim_{q \to 0} \lim_{\omega_n \to 0} (q^2/\pi^2) G(q,\omega_n) = K/(u \pi)$ (See also Appendix B in \cite{Sap1}). Hence, to solve this self-consistent equation, we assume that:
\begin{equation}
    G^{-1}_{\text{var}}(q,\omega_n) = \frac{1}{\pi K}\left(uq^2+\frac{\omega_n^2}{u} +\frac{F(\omega_n)}{u} \right)
    \label{eq:ansatz}
\end{equation}
where in the small $\omega_n$ limit, $F(\omega_n) = \eta(\alpha) \left| \omega_n \right|^{\psi_1}+a(\alpha)\left| \omega_n \right|^{\psi_2}$. We determine these parameters in the small $\omega_n$ limit.

\textit{Determination of $\psi_1$}: Using this form of the propagator, it can be easily seen that at large $\tau$ limit, one has $\int_{-\infty}^{\infty} dq' d\omega_{n'} G_{\text{var}}(q',\omega_{n'}) \left(1-\cos \omega_{n'}\tau \right) \approx C(\alpha) - \left(\frac{\zeta_{\tau}(\alpha)}{\tau}\right)^{1-\frac{\psi_1}{2}}$, where $C(\alpha)$ and $\zeta_{\tau}(\alpha)$ are $\alpha$-dependent constants. Using this, we obtain:
\begin{eqnarray}
&&\eta(\alpha) \left| \omega_n \right|^{\psi_1} + a(\alpha)\left| \omega_n \right|^{\psi_2} \stackrel{\text{large} \ \tau}{\approx} \nonumber\\
&&\int d\tau \frac{\left( 1 -\cos \omega_n \tau \right)}{\tau^{s+1}} \left( 1 +\left( \frac{\zeta_{\tau}(\alpha)}{\tau}\right)^{1-\frac{\psi_1}{2}}\right)
\end{eqnarray}
From power counting of both sides, we find out that $\psi_1=s$ and $\psi_2 = 1+\frac{s}{2}$. Note that $\psi_2$ is sub-leading for $s>0$.\\

\begin{figure*}[t!]
\centering
\includegraphics[width=1\linewidth, clip, trim=30 5 160 20]{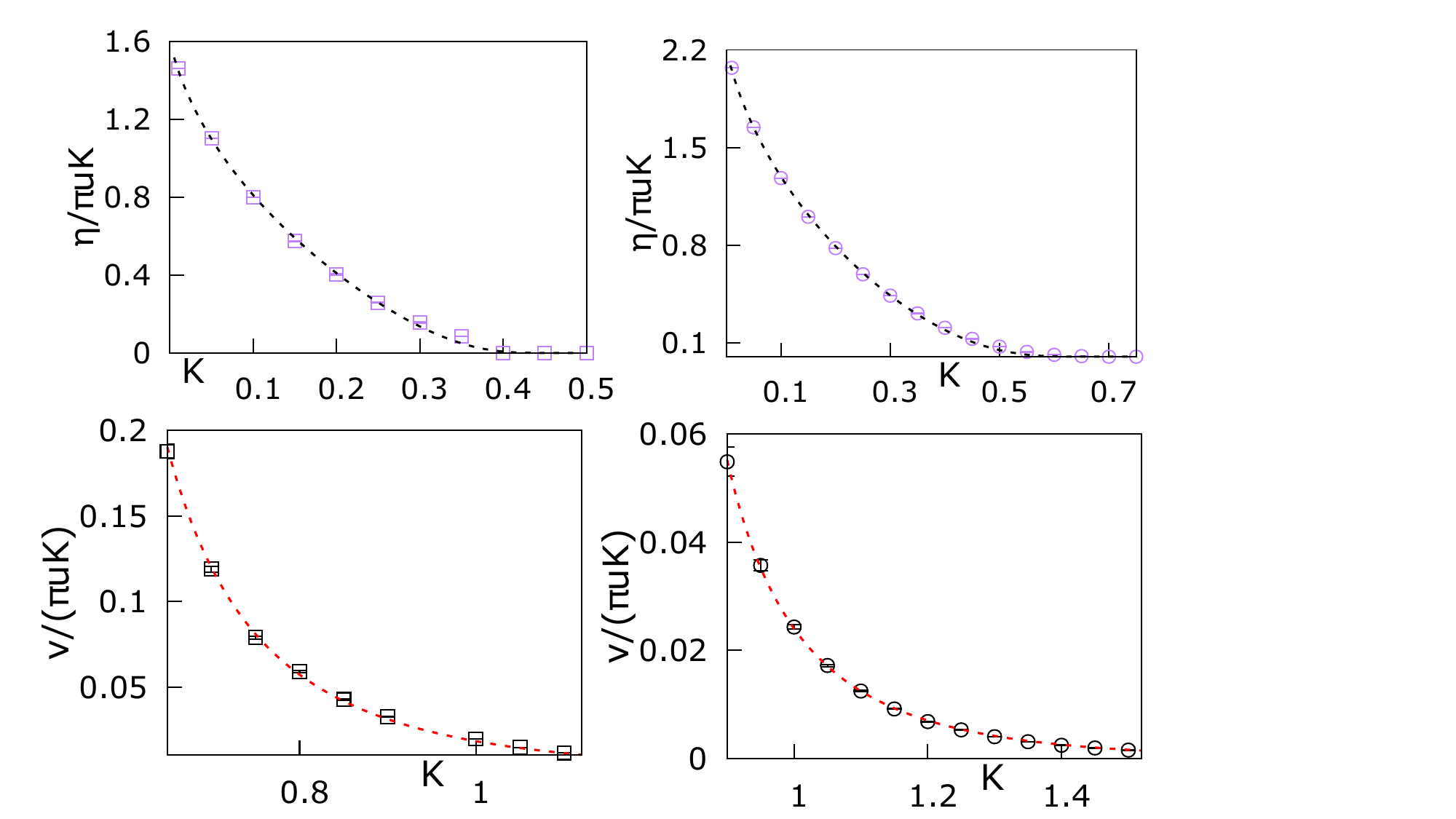}
\caption{
The parameters $\eta$ and $\nu$ obtained from the numerical solution of eq. (\ref{Eq_num}) (with $\beta=1024$ and $\alpha=5$). For $\eta$ (top row), we use $\alpha'$ and $\Lambda$ from eq. (\ref{Eq_eta_vs_K}) as  fitting parameters. For the ohmic case (purple square), $\alpha'= 10.096, \Lambda=1.963$, and for the subohmic case (purple circle), $\alpha'=8.29, \Lambda=3.29$. For the plot of $\nu$ (bottom row), the fitting parameters are $\tau_c$ and $\Lambda$ from eq. (\ref{Eq_omega2}). For the ohmic case (black square), $\tau_c=1.68, \Lambda=0.272$, and for the subohmic case (black circle), $\tau_c=1.241, \Lambda= 0.415$.}
\label{Fig_eta_vs_K}
\end{figure*}

\textit{Determination of $\eta$}: The behavior of the coefficient of $|\omega_n|^s$ ($\eta(\alpha)$) is important to locate the transition point between the LL and the dissipative phase. It can be estimated from the variational method. Indeed, neglecting the subleading term, we get $G^{-1}_{\text{var}}(q,\omega_n) = \frac{1}{ \pi K}\left(uq^2+\frac{\omega_n^2}{u}  + \frac{\eta}{u}|\omega_n|^s\right)$. Using this form of the propagator, it can be easily seen that $\int_{0}^{\infty} dq \int_{0}^{\Lambda} d\omega_n \ G_{\text{var}}(q,\omega_n) \approx \frac{2K}{2-s} \log \frac{4 \Lambda^{2-s}}{\eta}$, where $\Lambda$ is an ultraviolet cut-off. Plugging this result in eq. (\ref{eq:self_con}), we obtain:
\begin{equation}
\frac{\eta \omega_n^s}{uK} \stackrel{\text{small} \ \omega_n}{\approx} \alpha' \left(\frac{\eta}{\Lambda'^{2-s}}\right)^{\frac{2K}{2-s}} \omega_n^s
\end{equation}
Where $\alpha'$ depends on $\alpha, s$ and $\Lambda$, and $\Lambda'=4^{\frac{1}{2-s}}\Lambda$. Comparing the coefficient of $\omega_n^s$ on both sides, we see that there is a critical point at $K_c = 1-\frac{s}{2}$ where $\eta$ goes to zero. For $K<K_c$, the solution reads: 
\begin{equation}
\label{Eq_eta_vs_K}
\eta = 
    \left[\alpha' u K \Lambda'^{-2K} \right]^{\frac{2-s}{2-s-2K}}
\end{equation}

\subsection{LL phase}
To calculate the (eventual) renormalization of the coefficient of $\omega_n^2$ in the LL phase we consider that $F(\omega_n) = \nu \omega_n^2$. We assume that the correction coming from $\nu$ is small compared to $K$. Hence, to estimate $\nu$, we replace $G_{\text{var}}$ on the right side of the eq. (\ref{eq:self_con}) by the bare LL propagator $\pi K\left[uq^2 + \frac{\omega_n^2}{u} \right]^{-1}$. Hence, we find $\nu \omega^2_n=\frac{\alpha}{\pi^2} \int_{\tau_c}^{\infty} d\tau \frac{1-\cos \omega_n \tau}{\tau^{1+s}} \exp \left(-2K \int_0^{\Lambda} d\omega_n' \frac{1-\cos \omega_n' \tau}{\omega_n'} \right)$. The integral over $\omega_{n'}$ yields $(\gamma_E + \ln \Lambda \tau)$, and after expanding $\cos \omega_n \tau$ for small $\omega_n$, we find: 
\begin{equation}
\nu = \frac{\Tilde{\alpha}\exp(-2K\gamma)}{\tilde{\Lambda}^{2K}\left(2K+s-2\right)}
\label{Eq_omega2}
\end{equation}
Where $\Tilde{\alpha} = \frac{\alpha \tau_c^{2-s}}{2\pi^2}$ and $\Tilde{\Lambda}= \Lambda \tau_c$. We see that this estimate for $K>1-s/2$, large $\Lambda$ and small $\alpha$ represents a small correction to the action \footnote{For more details, see O. Bouverot-Dupuis, S. Majumdar, A. Rosso and, L. Foini, in preparation}. From the variational ansatz, we see that the Luttinger parameter $K$ is normalized to $K_r = K/\sqrt{1+\nu}$. This renormalization results from the fact that the variational procedure captures the perturbative renormalization group (RG) flow of $K$ up to the first order in $\alpha$. In the Sine-Gordon model, the variational solution does not renormalize the parameters $K, u$ and $\alpha$ \cite{giamarchibook}. Indeed, as we show in Appendix B, the perturbative RG flow of $K$ is non-zero even in the first order, which is also captured by the variational method.

\subsection{Numerical solution of the self-consistent equation}
To support our claim, we also numerically solved the following self-consistent equation for $F(\omega_n)$ by plugging eq. (\ref{eq:ansatz}) in eq. (\ref{eq:self_con}):
\begin{equation}
\label{Eq_num}
\begin{gathered}
     F(\omega_n) = \frac{uK\alpha}{\pi} \sum_{\tau=1}^{\beta-1}D(\tau) (1-\cos \omega_n \tau)\\
     \times \exp\left(-\frac{2\pi K}{\beta} \sum_{n'=-\frac{\beta}{2}}^{\frac{\beta}{2}-1} \frac{1-\cos \omega_{n'}\tau}{\sqrt{\omega_{n'}^2+F(\omega_{n'})}} \right) 
\end{gathered}
\end{equation} 
where $D(\tau)$ is the long-range kernel of eq. (\ref{eq:self_con}), realized on a discretized lattice with periodic boundary condition, namely $D(\tau)=\sum_{k=\beta/2}^{\beta/2-1} \mathcal{B} \left(\left(\tau+k\beta\right)-\frac{s}{2}, s-1 \right)$, where $\mathcal{B}()$ is the Beta-function (For more details, see App. C of \cite{Sap1}). In Fig. (\ref{Fig_F_omega}), we check the behavior of $F(\omega_n)$ for ohmic and subohmic baths in both LL and dissipative phases. Fig. (\ref{Fig_eta_vs_K}) shows us the behavior of $\eta$ and $\nu$ for dissipative phase and LL respectively for ohmic and subohmic baths. For fitting purposes, we use $\alpha'$ and $\Lambda$ for $\eta$ and $\tau_c$ and $\Lambda$ for $\nu$ as fitting parameters because they depend on the boundary condition and discretization. The plots show us that indeed our analytical predictions of Eq. (\ref{Eq_eta_vs_K}) and Eq. (\ref{Eq_omega2}) are in fair agreement with the direct numerical solution of Eq. (\ref{Eq_num}). 

\section{Order Parameter and Dissipative Phase}\label{Sec_OP}

In the dissipative phase, the spin chain develops a long-range order spin density wave. To better understand the properties of this phase we first study the order parameter of the transition, namely the amplitude of the SDW. Using Eq. (\ref{eq:sigma-boson}), together with the symmetry $\phi \to -\phi$  to remove the terms $\langle \nabla \phi \rangle$ and $\langle \sin(2\phi) \rangle$, we see:
\begin{equation}
    \langle \sigma^z(x) \rangle = \frac{1}{\pi a} \langle \cos(2\phi) \rangle \cos(2 q_F x)
\end{equation}
Comparing with eq. (\ref{eq:SDW}), we identify the amplitude of the SDW :
\begin{equation}\label{Order_parameter}
\sigma_1 = \frac{1}{\pi a} \langle \cos(2\phi(x,\tau))\rangle \ .
\end{equation}
We note two important points:
\begin{itemize}
    \item In contrast with the standard Peierls mechanism, the amplitude of the SDW is not associated with the formation of a gap. Indeed, the spin chain is gapless.\\

    \item For the incommensurate case the global shift $\phi \to  \phi+c$ does not cost any energy, but in the dissipative phase, this symmetry will be broken by the presence of local field or impurity. It is then convenient to fix this constant by setting the center of mass of the interface to zero, namely $\phi(q=0,\omega_n=0)=0$.
\end{itemize}. 

In the thermodynamic limit $L \to \infty$ and zero temperature limit $\beta\to \infty$, the order parameter is zero, in the LL phase (no true long-range order) whereas it is constant in the dissipative phase. Indeed, we can estimate the value of the order parameter in the dissipative phase, using the variational ansatz $G_{\text{var}}(q,\omega_n)=\pi K\left[ uq^2+ \eta \frac{\left |\omega_n\right|^s}{u}+\frac{\omega_n^2}{u} \right]^{-1}$: 
\begin{equation}
  \sigma_1=\frac{1}{\pi a}\langle \cos(2\phi) \rangle=\frac{1}{\pi a}e^{- \frac{2 }{\pi^2} \int\limits_0^{\Lambda}\text{d}\omega_n \int\limits_0^{\infty} dq \ G_{\text{var}}(q,\omega_n)}
    \label{eq:OP_def2}
\end{equation}

It is instructive to consider the effect of finite temperature and finite size. One can easily find out that in the Fourier space, the order parameter is given by $\langle \cos 2\phi \rangle_{L,\beta} = \exp \left( - \frac{2}{ \beta L} \sum\limits_{\substack{q,\omega_n\\ q,\omega_n \neq 0}} G_{\text{var}}(q,\omega_n) \right)$. \\
As shown in Appendix A, this sum can be decomposed into three contributions :
\begin{itemize}
    \item The contribution of $\omega_n=0, q \neq 0$ terms, which account for finite size effect.\\
    
     \item The contribution of $\omega_n \neq 0, q=0$ terms, which account for finite temperature effect.\\
     \item The contribution of $\omega_n \neq 0, q \neq 0$ terms, which can be approximated by eq. (\ref{eq:OP_def2}) with sub-leading corrections.
\end{itemize}

Using the variational action (eq. (\ref{eq:ansatz})) with the LL ansatz $F(\omega_n) = \nu \omega_n^2 $, one can find that (For details, see Appendix A):
\begin{equation}
  \langle \cos 2\phi \rangle_{L,\beta}^{\text{LL}} \sim e^{-\frac{\pi^2}{6} \left[\chi\frac{L}{\beta}+\rho_s \frac{\beta}{L}\right]-K_r \ln \min(\beta,L)}
   \label{eq:OP_LL_fin}
\end{equation}
Here $\chi$ is the susceptibility ($\pi \chi= K/u = K_r/u_r $) and $\rho_S= \lim_{q \to 0} \lim_{\omega_n \to 0} (\omega_n^2/\pi^2)G(q,\omega_n)$ is the spin stiffness ($\pi \rho_s= K_ru_r $).\\

Using the variational action with dissipative phase ansatz $F(\omega_n)=\eta |\omega_n|^s$, it behaves as:
\begin{equation}
      \langle \cos 2\phi \rangle_{L,\beta}^{\text{diss}}\sim \sigma_1 e^{-\chi\frac{\pi^2}{6} \frac{L}{\beta}+\frac{2 u K}{\eta}\frac{b_0(s)}{(2\pi)^{s-1}} \frac{\beta^{\kappa(s)}}{L}   +c_1 \beta^{\frac{s}{2}-1}}
      \label{eq:OP_diss_fin}
\end{equation}
Three limits should be discussed :
\begin{itemize}
    \item In the thermodynamic limit $L \to \infty$ and finite temperature, both order parameters vanish as $\sim \exp(-\pi^2 \chi L/6 \beta)$.\\ 
    \item In the zero temperature limit $\beta \to \infty$ and for a finite length $L$, in the LL regime, the order parameter $ \sigma_{1_{L,\infty}}^{\text{LL}}$ vanishes exponentially as $ \sim \exp(-\pi^2 \rho_s \beta/6 L)$. In the dissipative regime, the order parameter $ \sigma_{1_{L,\infty}}^{\text{diss}}$ vanish as a stretched exponential $\sim \exp(-\beta^{s-1}/L)$ for superohmic bath, while it converges to a constant in the subohmic case. This ordered phase at finite $L$ can be related to the transition observed for single particle models in the presence of a subohmic bath \cite{ThierryLedoussal1, ThierryLedoussal2}.\\
    \item In the numerical simulation, we set $L=\beta$ and send $\beta \to \infty$. In this limit, we find:
    \begin{eqnarray}
      \langle \cos 2\phi \rangle_{L=\beta=\infty}^{\text{LL}} &\sim& L^{-K_r}\\
    \langle \cos 2\phi \rangle_{L=\beta=\infty}^{\text{diss}} &\sim& \sigma_1 e^{-\frac{\pi^2 \chi}{6}}\left(1+c_1 L^{\frac{s}{2}-1} + c_2L^{s-2}\right) \nonumber   
    \end{eqnarray}
\end{itemize}

\subsection{Two-point correlation function}
 
To understand the nature of the order in the dissipative phase, it's important to introduce the two-point correlation functions:
\begin{eqnarray}
    \langle\sigma^z_{x,\tau}\sigma^z_{0,\tau} \rangle &\sim& \langle e^{i 2\phi(x,\tau)}e^{-i 2\phi(0,\tau)}\rangle \cos \left(2q_Fx\right)\nonumber \\
    \langle\sigma^z_{x,\tau}\sigma^z_{x,0} \rangle &\sim& \langle e^{i 2\phi(x,\tau)}e^{-i 2\phi(x,0)}\rangle
\end{eqnarray}
Note that The spatial spin-spin correlator has an overall oscillating factor of $\cos(2q_Fx)$, which doesn't affect the decay of the correlator at large $x$.  Under the gaussian variational approximation, one can see that $\langle e^{i 2\phi(x,\tau)}e^{-i 2\phi(0,\tau)}\rangle=e^{-2\langle\left(\phi(x,\tau)-\phi(0,\tau)\right)^2\rangle} \equiv e^{-2B(x)}$ and similarly for $\langle e^{i 2\phi(x,\tau)}e^{-i 2\phi(x,0)}\rangle = e^{-2B(\tau)}$. From eq. (\ref{eq:tau_r}) and eq. (\ref{eq:x_r}) of Appendix B, one can easily see that for large $x$ at finite temperature and in the thermodynamic limit :
\begin{align}
    \langle\sigma^z_{x,\tau}\sigma^z_{0,\tau} \rangle_{\text{LL}} &\sim \exp\left(-\frac{2\pi^2 \chi x}{\beta}\right)x^{-2K_r} \cos\left(2q_Fx\right) \nonumber \\
    \langle\sigma^z_{x,\tau}\sigma^z_{0,\tau} \rangle_{\text{diss}} &\sim \sigma_1^2 \exp\left(-\frac{2\pi^2 \chi x}{\beta}\right) \left(1+a_2 x^{1-\frac{2}{s}} \right) \nonumber\\
    &\times \cos (2 q_F x)
\end{align}
and for large $\tau$ at zero temperature and for finite $L$:
\begin{eqnarray}
    \langle\sigma^z_{x,\tau}\sigma^z_{x,0} \rangle_{\text{LL}} &\sim& \exp\left(-\frac{2\pi^2 \rho_s \tau}{L}\right)(u_r \tau)^{-2K_r} \\
    \langle\sigma^z_{x,\tau}\sigma^z_{x,0} \rangle_{\text{diss}} &\sim& \sigma_1^2\exp\left(-\frac{K u \tau^{f(s)}}{\eta L} \right)\left(1+a_1 \tau^{\frac{s}{2}-1} \right)\nonumber
\end{eqnarray}
Where $f(s) = 0$ for subohmic bath and $f(s) = 1-s$ for superohmic bath. 

These results show that in the limit of finite temperature, above a lengthscale $\beta/2 \pi^2 \chi$, both the order in the dissipative phase as well as the quasi-order in the LL phase are exponentially suppressed. On the other hand at $T=0$ there is long-range order:
\begin{equation}
\begin{array}{l}
\displaystyle
\lim_{x \to \infty}
\langle\sigma^z_{x,\tau}\sigma^z_{0,\tau} \rangle_{\text{diss}} =  \sigma_1^2 \cos \left( 2 q_F x\right)
\end{array}
\end{equation}

Connected spatial and imaginary time correlations decay in a power law fashion at $T=0$, with an exponent which increases upon decreasing $s$.  These results, along with the behavior of the order parameter, show that at zero temperature, the dissipative phase is indeed an SDW with a gapless spectrum and long-range order. This ordered phase exists due to the spontaneous breaking of the continuous symmetry $\phi \to \phi+c$ due to the presence of the long-range dissipative action $S_{\text{int}}$. 

\section{Numerical simulations}\label{Sec_NS}

\begin{figure*}[t!]
\centering
\includegraphics[width=1\linewidth, clip, trim=45 50 55 35]{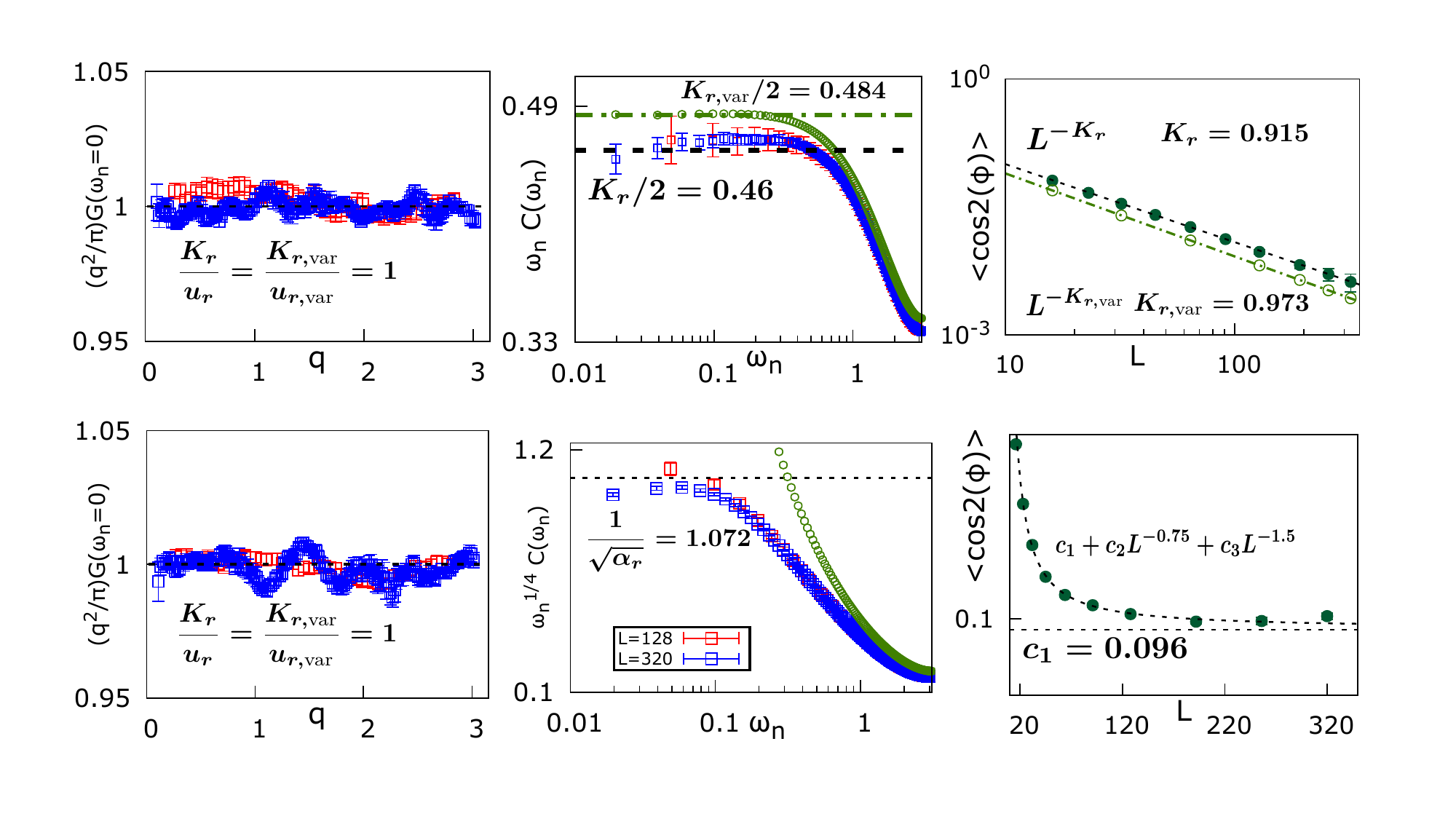}
\caption{Calculation of different quantities for $K=1$ that characterizes LL ($\alpha=2$, top row) and dissipative phase ($\alpha=6$, bottom row). Red points correspond to $L=\beta=128$ and blue points correspond to $L=\beta=320$. Green points correspond to $\omega_n C(\omega_n)$ calculated from numerically solving the self-consistent variational eq. (\ref{eq:self_con}). (\textit{left}) $\pi \chi$ remains unrenormalized for all values of $\alpha$ and $q$. (\textit{middle}) For $\alpha=2$, $\omega_n C(\omega_n)$ saturates to $K_r/2=0.46$ for small $\omega$, whereas $\omega_n^{0.25} C(\omega_n)$ saturates to $1/\alpha_r = 1.072$ for $\alpha=6$. The variational solution saturates to $K_{r,\text{var}}/2=0.486$ for $\alpha=2$ and fails to correctly predict the dissipative phase for $\alpha=6$. (\textit{right}) For $\alpha=2$, $\langle \cos \phi \rangle$ decays as a power law with the exponent $K_r=0.915 $. However, it saturates to a constant $c_1$ algebraically for $\alpha=6$. The fit for the order parameter in the dissipative phase gives $c_1=0.096,c_2=0.112$, and $c_3=3.37$. In the LL phase, $\langle \cos(2\phi)\rangle$ calculated with variational method decays with $K_{r,\text{var}}=0.973$.}
\label{fig:orderpar} 
\end{figure*}
\begin{figure}[h!]
\centering
\includegraphics[width=1\linewidth, clip, trim=5 10 15 20]{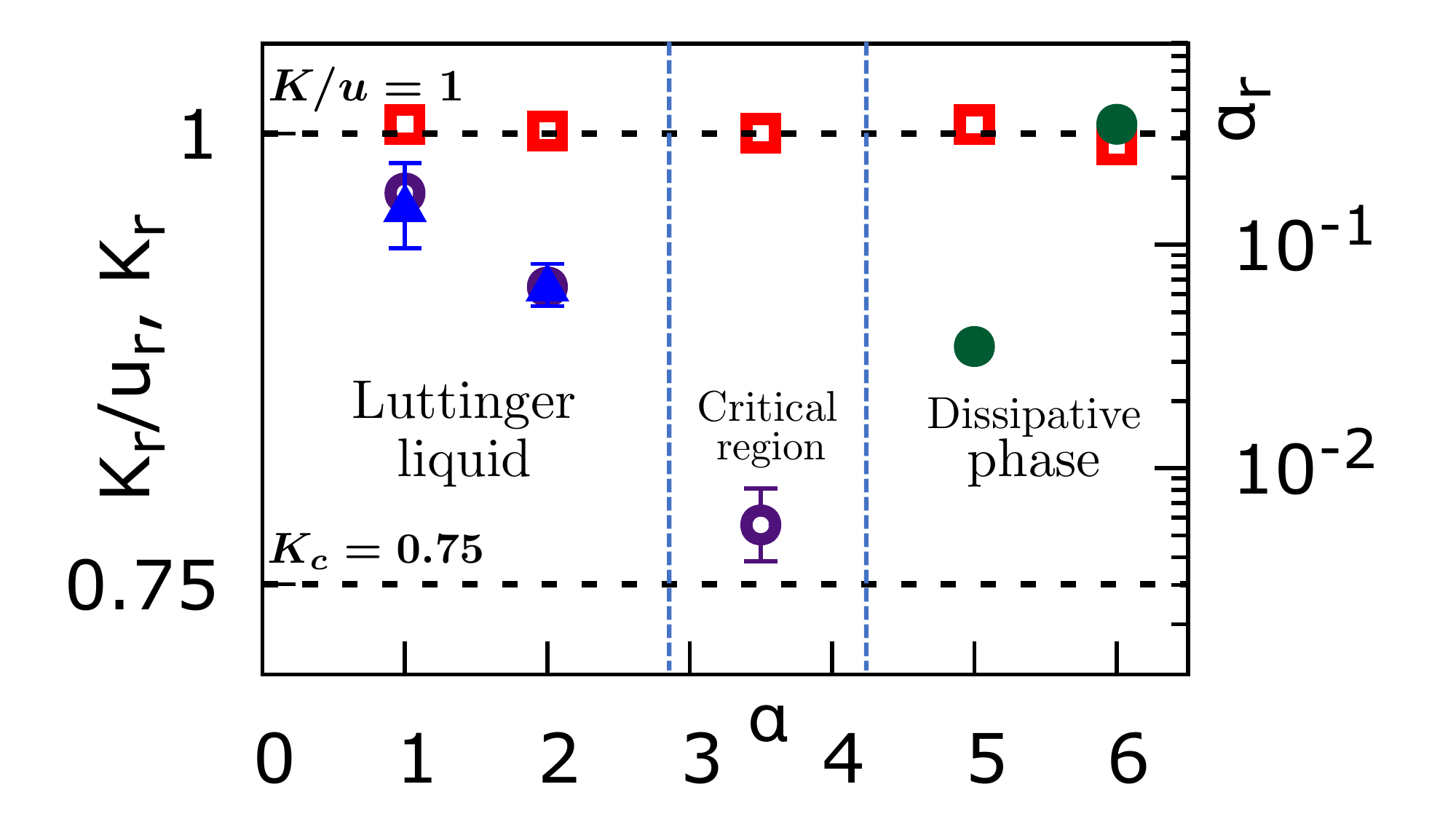}
\caption{Renormalized value of different parameters of the action for $K=1,s=0.5$. $K_r/u_r$ (red square) remains constant and equal to $1$ for all values of $\alpha$. $K_r$ decreases from $K=1$ to $K_c=0.75$ as $\alpha$ increases and approaches $\alpha_c$. $\alpha_r$ becomes relevant in the dissipative phase and increases as a function of $\alpha$. This behavior of the parameters helps us locate the critical region $\alpha_c \in (3,4)$.}
\label{fig:phase} 
\end{figure}

We verify the validity of our variational ansatz, both qualitatively and quantitatively, via numerical simulation of the original action with the cosine potential, Eq. (\ref{eq:action}). We numerically solve the Langevin dynamics differential equation associated with the action, namely the stochastical differential equation $\frac{d\phi(t)}{dt} = -\frac{\partial S_{\text{eff}}}{\partial \phi} + \Gamma(t)$, where $\Gamma(t)$ is Gaussian white noise with $\langle \Gamma(t) \rangle=0, \ \langle \Gamma(t)\Gamma(t') \rangle = 2\delta(t-t') $. Note that $\Gamma$ is the noise that thermalizes to $\exp(-S_{\text{eff}})$ and is not related to the temperature of the dissipative bath, which is zero. Discretizing the action and applying periodic boundary conditions in both $x$ and $\tau$ direction, we obtain the following differential equation that we simulate numerically:
\begin{equation}
    \begin{split}
        \frac{d\phi_{ij}(t)}{dt} &= \frac{1}{K \pi u}\left(\phi_{i+1,j}+\phi_{i+1,j}-2\phi_{i,j} \right)\\
        &+ \frac{u}{K\pi}\left(\phi_{i,j+1}+\phi_{i,j-1}-2\phi_{i,j} \right)+ \Gamma_{ij}(t)\\
        &+\frac{\alpha}{\pi^2} \sum_{i'}D(\left|i-i'\right|)\sin \left[2 \left(\phi_{i'j}-\phi_{ij} \right) \right]
        \label{eq:langevin}
    \end{split}
\end{equation}
Where $i \in (1,\beta)$ and $j \in (1,L)$ represents the discretized $\tau$ and $x$ indices respectively. We solve this differential equation at long time and obtain equilibrated configurations $\phi_{\text{eq}}(x,\tau)$. We then calculate various correlation functions on these configurations and match them against our analytical predictions. We compare the Langevin equation simulation with the variational method prediction, obtained from numerically solving Eq. (\ref{eq:self_con}). The values of the parameters chosen for both simulations are $K=1, u=1, s=0.5$ and $dt=0.05$, where $dt$ is the Langevin time-step. We varied the value of $\alpha$, and for each value of $\alpha$, we simulate Eq. (\ref{eq:langevin}) for different sizes, scaling $L=\beta$. From the variational study, we expect that there exists a critical dissipative strength $\alpha_c(K)$ such that for $\alpha<\alpha_c$, the correlation functions will correspond to the LL propagator $G^{-1}_{\text{LL}}=\frac{1}{\pi K}\left(u q^2+\frac{\omega_n^2}{u}(1+\nu) \right)$, and for $\alpha>\alpha_c$, they will behave according to the dissipative phase propagator $G^{-1}_{\text{var}} = \frac{1}{\pi K} \left( u q^2 + \frac{\eta |\omega_n|^s}{u}+\frac{a_1\left|\omega_n \right|^{1+\frac{s}{2}}}{u}+\frac{a_2 \omega_n^2}{u} \right)$. In Fig. \ref{fig:orderpar}, we show the results for $\alpha=2$ (top row), which we find to be in the LL phase, and $\alpha=6$, which turns out to be in the dissipative phase. The first quantity we compute is $(q^2/\pi) G(q,\omega_n)$. Fig. \ref{fig:orderpar}, left, shows that this quantity, both with the Langevin method and the variational method, remains unrenormalized and equal to $K/u$ for all values of $q$ and both values of $\alpha$. This is in agreement with our variational ansatz. Next, we compute $C(\omega_n) = \frac{1}{\pi L} \sum \limits_q G(q,\omega_n) $. This quantity is useful for extracting and differentiating between the $\omega_n$ dependence of $G(q,\omega_n)$ in the two phases. Indeed, for small $\omega_n$, $C(\omega_n)$ behaves as:
\begin{equation}
   C(\omega_n \to 0) = 
   \begin{cases}
       \frac{K_r}{2 \omega_n}, \ & \mbox{LL}\\
       \frac{1}{\sqrt{\alpha_r}\omega_n^s}\ & \mbox{dissipative}
   \end{cases} 
\end{equation}
where $K_r = \frac{K}{\sqrt{\nu+1}} $ and $\alpha_r=4\eta/K^2$. We denote the renormalized value of $K$ obtained from the Langevin simulation as $K_r$ and the numerical variational solution as $K_{r,\text{var}}$.  Fig. \ref{fig:orderpar}, middle, shows that indeed for $\alpha=2$, $\omega_n C(\omega_n)$ saturates to a constant, whereas for $\alpha=6$, $\omega_n^{0.25} C(\omega_n)$ goes to a constant as $\omega_n \to 0$, indicating that $\alpha=2$ is in LL phase and $\alpha=6$ is in the dissipative phase. The variational solution also shows a renormalization of $K$, for example for $\alpha=2$, we get $K_{r,\text{var}}=0.968$. This result is in fair agreement with the Langevin simulation, $K_r=0.92$. However, at large $\alpha$, the variational method fails and estimates the transition at $\alpha_c=10$. From fig. \ref{fig:orderpar}, middle bottom, for $\alpha=6$ the system is already in the dissipative phase. For our third and final check, we show the behavior of the order parameter (eq. (\ref{Order_parameter})). To extrapolate to the zero temperature behavior, we compute $\langle \cos(2\left(\phi-\phi_{\text{CoM}}\right)) \rangle$. Fig. (\ref{fig:orderpar}), left, shows that this quantity decays as a power law of the system size for $\alpha=2$ (top) and saturates to a constant for $\alpha=6$ (bottom). Therefore, we confirm the existence of a phase transition between LL and a new dissipative phase induced by the bath. This new phase has unaltered susceptibility, gapless spectrum, and vanishing spin stiffness $\rho_s$. In Fig. \ref{fig:phase}, we show the renormalized values of different parameters as a function of $\alpha$, which tells us that for $K=1$, $\alpha_c \in (3,4)$. 

\section{Conductivity}\label{Sec_Cond}

From Linear Response theory, the conductivity can be determined via the analytic continuation of the propagator \cite{giamarchibook}:
\begin{equation}
    \sigma(\omega) = \frac{e^2}{\pi^2 \hbar} \left[ \omega_n G(q=0,\omega_n)\right]_{i\omega_n \to \omega+i\epsilon }
\end{equation}
Where $\epsilon$ is a small positive number close to zero. Using our ansatz we find that the DC conductivity $\sigma_{\text{DC}} \equiv \text{Re}(\sigma(\omega \to 0)) = \lim_{\epsilon \to 0}(e^2/\pi^2 \hbar) \epsilon^{1-s}$, which goes to zero for subohmic ($s<1$) baths. This supports our claim that the system for a subohmic bath in the dissipative phase is insulating at zero temperature.

\section{Conclusions}\label{Sec_Concl}

In this work, exploiting the bosonization formalism, we have shown via analytical and numerical methods that an incommensurate XXZ spin chain coupled to local baths undergoes an LL-dissipative phase transition at $T=0$. At transition, the chain undergoes a spontaneous symmetry breaking,  with an order parameter $\langle  \cos(2\phi) \rangle$,  that identifies with the  amplitude of a long-range  ordered  spin density wave. Remarkably, the spin wave is gapless and the order originates from the fractional nature of the excitations of the dissipative phase. Moreover, from the linear response, we observe a suppression of the DC conductivity that vanishes  for subohmic baths. Hence, it is tempting to compare this dissipative transition with the localization transition observed for quenched disorder \cite{GiamarchiSchulz1,GiamarchiSchulz2, Laflorencie}. There, the localized phase is also gapless and the fluctuations along the imaginary time direction are suppressed. However, the order parameter $\langle  \cos(2\phi) \rangle$ is zero (as there is no spontaneous breaking of a continuous symmetry) and the spatial spin-spin correlations decay to zero exponentially above a finite localization length.
in the dissipative phase instead, the spin-spin correlations decay  to a finite value with an $s$-dependent power law.
For slower baths (small $s$), the decay becomes faster, and the exponent  diverges in the limit $s\to 0$, signaling that (connected) correlations can decay exponentially.

In the future, we would like to study the properties of the model at finite temperatures by variational methods and numerical simulations. This would be very interesting in view of our interpretation of the bath as annealed disorder and this study could possibly shed some light on the ongoing discussion on the many-body localization transition.

Another direction that we we have taken is the study of the same model at half-filling. This was partially done in \cite{Cazalillalong} and we plan to do it in full generality.\\

\begin{acknowledgments}
{\it Acknowledgments:} This work was supported in part by the Swiss National Science Foundation under grant 200020-188687. This work was performed using HPC/AI resources from GENCI-IDRIS (Grant 2022-AD011013581) and GENCI-TGCC (Grant 2022-AD011013555). We thank Thibaud Maimbourg and Oscar Bouverot-Dupuis for their useful discussions. 
\end{acknowledgments}

\appendix

\section{System size dependence of Order Parameter}

In this section, we compute $\langle \cos 2\phi \rangle_{L,\beta} = \exp \left( - \frac{2}{ \beta L} \sum\limits_{\substack{q,\omega_n\\ q=\omega_n \neq 0}} G_{\text{var}}(q,\omega_n) \right)$. This sum can be decomposed into three terms:
\begin{eqnarray}
    S_1 &=& \frac{2}{\beta L}\left[\sum\limits_{q\neq 0}G_{\text{var}}(q,0) + \sum\limits_{\omega_n\neq 0}G_{\text{var}}(0,\omega_n)  \right. \nonumber\\
    &+& \left.\sum\limits_{q \neq 0,\omega_n \neq 0}G_{\text{var}}(q,\omega_n)\right]
    \label{eq:OP_sum}
\end{eqnarray}
The $q \neq 0, \omega_n \neq 0$ contributions can be converted as $\frac{1}{\beta L}\sum\limits_{q \neq 0,\omega_n \neq 0} \to \frac{1}{\pi^2}\int\limits_{1/L}dq\int\limits_{1/\beta}d\omega_n$. Using the eq. (\ref{eq:ansatz}) with $F(\omega_n)=\nu \omega_n^2$, we see that:
\begin{align}
\begin{split}
     \frac{2}{\beta L}\sum\limits_{q\neq 0}G(q,0)&=\frac{4\pi K}{u \beta L}\sum\limits_{m=1}^{\infty}\frac{1}{\left(\frac{2\pi m}{L}\right)^2} \nonumber \\
    &=\frac{\pi K L}{6 u \beta} \nonumber \\ 
    \frac{2}{\beta L} \sum\limits_{\omega_n\neq 0}G(0,\omega_n)&= \frac{4\pi K u}{\left(1+\nu\right)\beta L}\sum\limits_{n=1}^{\infty}\frac{1}{\left(\frac{2\pi n}{\beta}\right)^2} \nonumber\\
    &=\frac{\pi u  K \beta}{6(1+\nu)L}\\
   \frac{2}{\beta L} \sum\limits_{q \neq 0,\omega_n \neq 0}G_{\text{var}}(q,\omega_n) &=\frac{2K}{\pi}\int\limits_{1/L}^{\Lambda_1}\int\limits_{1/\beta}^{\Lambda_2} \frac{d\omega_n dq}{uq^2 + \frac{\omega_n^2}{u}\left(1+\nu \right)} \nonumber\\
   &\sim \frac{K}{\sqrt{1+\eta}} \ln \min(\beta,L) \nonumber
\end{split}   
\end{align} 
Similarly, with $F(\omega_n) = \eta |\omega_n|^s$, we find that the contribution from the first term is the same. The contribution from the second term can be written as:
\begin{equation}
   \frac{2}{\beta L} \sum\limits_{\omega_n\neq 0}G(0,\omega_n) = \frac{2 u K}{\eta}\frac{b_0(s)}{(2\pi)^{s-1}} \frac{\beta^{\kappa(s)}}{L}  
\end{equation}
Where $\kappa(s) = 0$ and $b_0(s) \sim \frac{1}{1-s}$for a subohmic bath ($0<s<1$),  $\kappa(s) = s-1$ and $b_0(s) \sim \zeta(s)$ for a superohmic bath ($1<s<2$). The ohmic case ($s=1$) is special and $b_0(s) \beta^{\kappa(s)}$ should be replaced with $\ln \beta + \gamma_E$. The contribution from the third term is given by:
\begin{eqnarray}
    \frac{2}{\beta L}\sum\limits_{q \neq 0,\omega_n \neq 0}G(q,\omega_n) &=& \frac{2 K}{\pi}\int\limits_{1/L}^{\infty}\int\limits_{1/\beta}^{\Lambda} \frac{d\omega_n dq}{uq^2 + \frac{\omega_n^2}{u}+\frac{\eta \omega_n^s}{u}} \nonumber\\
    &\sim& c_0 - c_1 \beta^{\frac{s}{2}-1}
    \label{eq:OP_diss}
   \end{eqnarray} 
Where $c_0$ and $c_1$ are positive constants that depend on $K,u,\eta,s$, and ultra-violet cut-off $\Lambda$. Putting these terms together, we find eq. (\ref{eq:OP_LL_fin}) and $(\ref{eq:OP_diss_fin})$.

\section{Roughness of $\phi(x,\tau)$ in the dissipative phase}
At zero temperature, in the Luttinger liquid phase, the field $\phi(x,\tau)$ grows logarithmically in both directions $x$ and $\tau$. Here we characterize the roughness of the field  $\phi(x,\tau)$ in the dissipative phase. In particular, We compute the following correlation functions:
\begin{eqnarray}
    B(\tau) \equiv \langle \left[\phi(x,0)-\phi(x,\tau)\right]^2 \rangle \label{Btau} \\
    B(x) \equiv \langle \left[\phi(x,\tau)-\phi(0,\tau)\right]^2 \rangle \label{Bx}
\end{eqnarray}
 In the dissipative phase, using eq. (\ref{eq:ansatz}) with $F(\omega_n)=\nu |\omega_n|^s$, eq. (\ref{Btau}) can be written in the Fourier space as
 \begin{equation}
      B(\tau) = \frac{2}{\beta L} \sum_{q,\omega_n} \left(1- \cos \omega_n \tau \right) G_{\text{var}}(q,\omega_n)
 \end{equation}
 The $\omega_n = 0$ terms vanish due to the presence of the cosine term in the numerator. Hence, we write the contributions from the terms $q=0,\omega_n \neq 0$ and $q \neq 0, \omega_n \neq 0$ separately:
 \begin{eqnarray}
     B(\tau) &=& \frac{2}{\beta L} \sum\limits_{\omega_n \neq 0}(1-\cos \omega_n \tau)G_{\text{var}}(0,\omega_n) \label{eq:B_tau_first} \\
     &+& \frac{2}{\beta L}\sum\limits_{q \neq 0, \omega \neq 0}(1-\cos \omega_n \tau)G_{\text{var}}(q,\omega_n) \nonumber
 \end{eqnarray}
 The summation of the first term on the RHS of eq. (\ref{eq:B_tau_first}) gives:
 \begin{eqnarray}
          \frac{2}{\beta L} &\sum\limits_{\omega_n \neq 0}&(1-\cos \omega_n \tau)G_{\text{var}}(0,\omega_n) \nonumber\\
          &=&\frac{4 \pi K u}{\beta L}\sum\limits_{n=1}^{\infty}\frac{1-\cos \omega_n \tau}{\omega_n^2 + \eta \omega_n^s} \\
          &\sim& \frac{K u}{\eta}\frac{\tau^{f(s)}}{L}\nonumber
\end{eqnarray}
Where $f(s) = 0$ for subohmic bath ($0<s<1$) and $f(s) = 1-s$ for superohmic bath ($1<s<2$). For the ohmic case ($s=1$), $\tau^{f(s)}$ should be replaced by $\ln \tau$. For the second term on the RHS of eq. (\ref{eq:B_tau_first}), we convert the sum $\frac{1}{\beta L}\sum\limits_{q \neq 0, \omega_n \neq 0} \to \frac{1}{\pi^2}\int dq d\omega_n$ to find:
\begin{eqnarray}
          \frac{2}{\beta L}&\sum\limits_{q \neq 0, \omega \neq 0}&(1-\cos \omega_n \tau)G_{\text{var}}(0,\omega_n) \\
          &=& \frac{2K}{\pi} \int_0^{\Lambda} d\omega_n \int_0^{\infty} dq \frac{(1-\cos \omega_n \tau)}{uq^2 + \frac{\omega_n^2}{u} + \frac{\eta \omega_n^s}{u}} \nonumber
 \end{eqnarray}
  
The integral over $1$ in eq. (\ref{eq:tau_r}) gives us the same constant $c_0$ from eq. (\ref{eq:OP_diss}). The integral over $\cos \omega_n \tau$ gives us the $\tau$ dependence of $B(\tau)$, and we see that for large $\tau$ :
\begin{equation}
    B(\tau) \sim \frac{K u}{\eta} \frac{\tau^{f(s)}}{L}+c_0 - a_1 \tau^{\frac{s}{2}-1} 
    \label{eq:tau_r}
\end{equation}
Where $a_1=\frac{K}{\eta} \ \Gamma(1-\frac{s}{2}) \sin \left(\frac{\pi s}{4} \right)$.\\ 

Similarly, eq. (\ref{Bx}) can be written in the Fourier space and calculated :
\begin{eqnarray}
     B(x) &=& \frac{2}{\beta L} \sum\limits_{q \neq 0} \left(1- \cos qx \right) G_{\text{var}}(q,0) \\
     &+&\frac{2}{\beta L} \sum\limits_{q,\omega_n} \left(1- \cos qx \right) G_{\text{var}}(q,\omega_n) \nonumber
\end{eqnarray}
Like $B(\tau)$, we compute $B(x)$ termwise:
\begin{eqnarray}
             \frac{2}{\beta L} &\sum\limits_{q \neq 0}& \left(1- \cos qx \right) G_{\text{var}}(q,0)\\
             &=&\frac{4 \pi \chi}{\beta L}\sum\limits_{n=1}^{\infty}\frac{1-\cos qx}{q^2} \sim \frac{\pi^2 \chi x}{\beta} \nonumber \\
             \frac{2}{\beta L} &\sum\limits_{q,\omega_n}& \left(1- \cos qx \right) G_{\text{var}}(q,\omega_n)\\
             &=&\frac{2K}{\pi}\int_0^{\infty} d\omega_n \int_0^{\Lambda} dq \frac{(1-\cos qx)}{uq^2 + \frac{\omega_n^2}{u} + \frac{\eta \omega_n^s}{u}} \nonumber\\
             &\sim& c_0 - a_2 x^{1-\frac{2}{s}} \nonumber
 \end{eqnarray}
 \vspace{1cm}
 where $a_2=\frac{2K u^{\frac{2}{s}-1}\eta^{-\frac{1}{s}}\Gamma \left( \frac{2}{s}-1 \right)}{s}$. 
 Putting all the terms together, we obtain that for large-$x$:
\begin{equation}
  B(x) \approx\frac{\pi^2 \chi x}{\beta}+ c_0 - a_2 x^{1-\frac{2}{s}}
   \label{eq:x_r}
\end{equation}

In conclusion, in the thermodynamic limit where $L \to \infty$, the interface is flat in the $\tau$ direction. Along the $ x$ direction, it is rough at finite temperature and becomes flat at zero temperature. In this limit, both $B(x)$ and $B(\tau)$ algebraically saturate to the same constant but with different power laws, showing that there is long-range order in this phase.\\

\section{RG calculation}
In this section, we systematically derive the RG flow equations of the LL parameter K and the coupling strength $\alpha$. To analyze the RG flow of the parameters, we calculate the following correlation function :
\begin{equation}
R(r_1 - r_2) = \langle e^{i a \phi(r_1)}e^{-i a \phi(r_2)} \rangle, \ r = (x,u\tau)
\end{equation}
We know that for the quadratic LL action, $R(r_1-r_2) \sim \left(\frac{ r_1-r_2}{b} \right)^{-a^2K/2} $, where $b$ is some short-scale length cut-off. We perturbatively expand the correlation function with respect to $S_{\text{diss}} $. The perturbative series up to first order of $\alpha$ is given by $\langle e^{i a \phi(r_1)}e^{-i a \phi(r_2)} \rangle_{S_0} + \langle e^{i a \phi(r_1)}e^{-i a \phi(r_2)} \rangle_{S_0} \langle S_{int} \rangle_{S_0} - \langle e^{i a \phi(r_1)}e^{-i a \phi(r_2)} S_{int} \rangle_{S_0}$. The $0$th order term can be easily computed and is given by $\exp(\frac{-a^2 K}{2}F(r_1-r_2)),F(r) = \frac{1}{2} \log \left[ \frac{x^2 + (u |\tau|+b)^2}{b^2}\right]$. After computing the first-order contribution, we obtain:
\begin{widetext}
  \begin{eqnarray}
      && R(r_1-r_2) = e^{\frac{-a^2 K}{2}F(r_1-r_2)}[1+ \nonumber\\
    &&\frac{\alpha}{2 \pi bu^2} \int d^2r' d^2r'' e^{-2KF(x'-x',\tau'-\tau'')}B\sum_{\epsilon = + -} \left[e^{ aK\epsilon(F(r_1-r')-F(r_1-r'')-F(r_2-r')+F(r_2-r''))} -1 \right]]
  \end{eqnarray}   
\end{widetext}
where $B = \delta(x'-x'') D(\tau'-\tau'')$. After transforming the equation into CoM $R = \frac{r'+r''}{2}$ and relative co-ordinates $r = r' - r''$ and taylor expanding $F$ for small $r$, we expand the exponential for small value of $r$:
\begin{eqnarray}
&&R(r_1-r_2) = e^{\frac{-a^2 K}{2}F(r_1-r_2)}[1 + \nonumber\\
&&\frac{\alpha a^2K^2}{2 \pi bu^2} \int d^2r d^2R e^{-2KF(r)} B\nonumber\\ 
&&\times  (r.\nabla_R[F(r_1-R)-F(r_2-R)])^2
\end{eqnarray}
The term inside the square produces terms like 
$r_ir_j(\nabla_{R_i}[F(r_1-R)-F(r_2-R)]))(\nabla_{R_j}[F(r_1-R)-F(r_2-R)]))$, where $i,j$ denotes the two possible coordinates $x, y=u\tau$.  For the integral over $d^2r$ and by symmetry $x \to -x, y \to -y$, only the diagonal $i=j$ terms survive. The action is anisotropic whose effect can be included with an additional term in F of the form $ d \cos (2\theta)$, where $\theta$ is the angle between vector $(x,u\tau)$ and $x$ axis and $d$ is the measure of anisotropy.  After expanding the gradient terms and integrating by parts over $R$, we obtain two terms $I_{\pm}=\int d^2 R [F(r_1-R)-F(r_2-R)](\partial^2_X \pm \partial^2_Y) [F(r_1-R)-F(r_2-R)]$. The $I_+$ term renormalizes $K$ and $\alpha$, whereas the other term renormalizes the anisotropy which we are not interested in. Hence,
\begin{widetext}
    \begin{eqnarray}
     &&R(r_1-r_2) =e^{\frac{-a^2 K}{2}F(r_1-r_2)} \times\nonumber \\
     &&[1 - \frac{\alpha a^2 K^2}{4 \pi bu^2} \int d^2r d^2R e^{-2KF(r)} r^2 B [F(r_1-R)-F(r_2-R)](\nabla^2_X+\nabla^2_Y)[F(r_1-R)-F(r_2-R)] 
   \end{eqnarray}
\end{widetext}
As $F$ is a logarithmic function, we know that $ (\nabla^2_X+\nabla^2_Y)F(R) = 2 \pi \delta(R)$. After re-exponentiating the term inside the bracket, we obtain :
\begin{equation}
K_{\text{eff}} = K - \frac{2 \alpha K^2}{b u^2} \int_{r>b} d^2r r^2 \exp(-2KF(r)) B
\end{equation}

To understand the scaling of $K$ and $\alpha$, we express $d^2r$ and $r^2$ in terms of $x, u\tau$ and compute the integral over $\delta(x)$. Noticing that $ D(\tau)$ cancels out $\tau^2$,  we find:
\begin{align}
\begin{split}
K_{\text{eff}} = K - 2\alpha K^2 \int_{b}^{\infty} \frac{dy}{b}(\frac{y}{b})^{-2K}, \ y = u\tau
\end{split}
\end{align}

Sending $b$ to $b'=b+db$, we find :
\begin{eqnarray}
&K_{\text{eff}} = K - 2\alpha K^2\frac{db}{b}-2\alpha K^2 \int_{b'}^{\infty} \frac{dy}{b}(\frac{y}{b})^{-2K}\nonumber \\
&\implies K(b') = K(b)- 2\alpha(b) K^2(b) \frac{db}{b}
\end{eqnarray}
Similarly, 
\begin{equation}
\alpha(b') = \alpha(b)\left(\frac{b'}{b}\right)^{1-2K}
\end{equation}

If we parametrize $b = b_0e^l $, we obtain the following flow equations:
\begin{equation}\label{Eq_flow}
\begin{array}{l}
\displaystyle
\frac{dK}{dl} = -2\alpha K^2
\\ \vspace{-0.2cm} \\
\displaystyle
\frac{d\alpha}{dl} = (2-s-2K)\alpha
\end{array}
\end{equation}
These equations indicate the existence of a critical point $K_c = 1-\frac{s}{2}$, as in \cite{cazaillashort}, but the precise value of the numerical coefficient in the first equation of (\ref{Eq_flow}) differs of a factor of 2.\\
\bibliography{references_new}

\end{document}